\documentclass[a4paper,aps,pre,floatfix,twocolumn, superscriptaddress,nobibnotes,nolongbibliography,nofootinbib]{revtex4-2}

\usepackage[pdftex]{graphicx}
\setkeys{Gin}{width=0.9\columnwidth}

\usepackage{times}
\usepackage{verbatim}
\usepackage{bbold}
\usepackage{bbm}
\usepackage{bm}
\usepackage{lipsum}

\usepackage{amsmath}
\usepackage{amsfonts}
\usepackage{color}
\usepackage[english]{babel}
\usepackage{microtype}
\usepackage{comment}
\usepackage[normalem]{ulem}
\usepackage{float}

\usepackage{soul,xcolor}
\setstcolor{red}

\usepackage{fancyhdr}
\usepackage{relsize}

\usepackage[hidelinks,unicode=true]{hyperref}
\hypersetup{
colorlinks=true,
linkcolor=blue,
urlcolor=blue,
citecolor=blue,
pdfhighlight=/N}

\usepackage{comment}
\usepackage{lineno}

\usepackage{relsize}

\newcommand{\NoteR}[1]{{\color{red} (RPS: {#1})}}
\newcommand{\NoteF}[1]{{\color{orange} (FR: {#1})}}

\newcommand{\av}[1]{\langle #1 \rangle}

\begin{document}

\title{Modeling individual attention dynamics on online social media}

\author{Jaume Ojer}
\affiliation{Departament de F\'isica, Universitat Polit\`ecnica de Catalunya, Campus Nord, 08034 Barcelona, Spain}

\author{Filippo Radicchi}
\affiliation{Center for Complex Networks and Systems Research, Luddy School of Informatics, Computing, and Engineering, Indiana University, Bloomington, Indiana 47408, USA}

\author{Santo Fortunato}
\affiliation{Center for Complex Networks and Systems Research, Luddy School of Informatics, Computing, and Engineering, Indiana University, Bloomington, Indiana 47408, USA}

\author{Michele Starnini}
\affiliation{Department of Engineering, Universitat Pompeu Fabra, 08018 Barcelona, Spain}
\affiliation{CENTAI Institute, 10138 Turin, Italy}

\author{Romualdo Pastor-Satorras}
\affiliation{Departament de F\'isica, Universitat Polit\`ecnica de Catalunya, Campus Nord, 08034 Barcelona, Spain}

\date{\today}

\begin{abstract}
In the attention economy, understanding how individuals manage limited attention is critical. 
We introduce a simple model describing the decay of a user's engagement when facing multiple inputs. We analytically show that individual attention decay is determined by the overall duration of interactions, not their number or user activity. 
Our model is validated using data from Reddit's Change My View subreddit, where the user's attention dynamics is explicitly traceable. Despite its simplicity, our model offers a crucial microscopic perspective complementing macroscopic studies.
\end{abstract}

\maketitle


In an age characterized by an overwhelming abundance of information, the concept of the attention economy has emerged as a central framework for understanding how individuals navigate the digital landscape~\cite{huberman07, huberman13, bueno16, menczer20}. 
Within this framework, attention, rather than information, is the truly scarce resource: Understanding how attention is captured, distributed, and ultimately decays, is crucial~\cite{wu07, hodas_attention_2013, ciampaglia_production_2015, parolo15, li_collective_2022}. 
Online social media platforms, for instance, are designed to facilitate interaction and information sharing, but bomb users with a constant stream of stimuli, including posts, comments, notifications, and messages~\cite{rodriguez_quantifying_2014}. 
Most of the previous research has focused on patterns of collective attention within these environments, 
such as the emergence of trending topics~\cite{asur_trends_2011, aiello_sensing_2013, collective_wiki}, information cascades~\cite{sreenivasan_information_2017, gleeson_branching_2021}, and the spread of memes~\cite{weng_competition_2012, gleeson_competition-induced_2014}.

However, individual-level attention dynamics, specifically how a single user's focus evolves in response to multiple stimuli, remains mostly unexplored. 
This paper addresses this gap by introducing a simple model that describes the temporal decay of an individual's attention when confronted with a multitude of interactive inputs. 
For example, consider a user who posts a thought-provoking question on a social media platform and receives numerous replies from different users, each demanding a piece of their attention~\cite{vaca_ruiz_modeling_2014, feng_competing_2015}. 
Similarly, the model could describe the attention dynamics in email inbox management: Individuals receive a stream of emails, each varying in importance, competing for limited attention~\cite{whittaker_email_1996, barabasi_origin_2005, wainer_should_2011}. 
Likewise, in a collaborative work environment, a professional might receive numerous task assignments, comment threads, and status updates, and have to decide which assignment to prioritize~\cite{ludlow_decisions_2021, bugayenko_prioritizing_2023}. 
In all these examples, the model provides a framework to understand how an individual's finite attention is distributed and how it decays when faced with a large number of inputs.

To test our model, we focus on Reddit, a prominent online social news aggregation, content rating, and discussion platform. 
Reddit is organized into thousands of user-created communities known as subreddits, each dedicated to a specific topic. We specifically choose the subreddit \textit{Change My View} (CMV) as our empirical testbed. 
Unlike many other subreddits, where users might post and then disengage, CMV is unique in its explicit goal: Users post an opinion and invite others to challenge it, actively encouraging and valuing engagement with the comments. 
Previous studies have focused on processing the text of comments in CMV, investigating the effects that the language structure can have on the power of persuasion in the discourse~\cite{tan_winning_2016, musi_changemyview_2018, mensah_characterizing_2019, dutta_changing_2020} and on the process of reaching consensus~\cite{monti_language_2022}. 
Furthermore, CMV has been widely used to study users' opinion change on social media~\cite{tan_winning_2016, mensah_characterizing_2019, monti_language_2022}, among other online platforms~\cite{xiong_opinion_2014}. 
Crucially, on CMV, users explicitly acknowledge comments that successfully ``change their view", allowing us to observe and quantify how individual attention evolves in response to various inputs. 
This inherent interest in following and responding to discussions makes CMV an ideal environment for studying individual attention decay, as users are demonstrably invested in the interactions surrounding their posts.

Our model aims to predict the probability that an individual will allocate their attention to specific interactions from others, over time. 
The model assumes that the focal individual rewards their peers with attention simply based on their activity in writing comments. 
We derive an analytical solution for the attention decay as a function of the number of interactions received, their cumulative duration, and the distribution of activity among the interacting individuals. 
We found that the individual attention dynamics is fully determined by the overall duration of interactions, while independent of the number of interactions received and the users' activity. 
Despite its simplicity, the model is able to capture the individual attention decay of users on CMV. 
Moreover, the model reproduces the interevent time between consecutive rewards, from both the point of view of the original poster (OP) and comments' authors (or commenters). 
By modeling individual attention dynamics, we provide a microscopic perspective that complements the macroscopic view offered by collective attention studies~\cite{lehmann_dynamical_2012, sasahara_quantifying_2013, eom_twitter-based_2015, mocanu_collective_2015, bao_uncovering_2017, garimella_effect_2017, he_measuring_2017, lorenz-spreen_accelerating_2019, de_domenico_unraveling_2020, bento_evidence_2020, kobayashi_modeling_2021}.

\emph{Model definition}. 
We consider a thread, consisting of $N$ total comments and a duration of $T$ days. We denote by $n(t)$ the number of comments generated when $t$ days have elapsed since the beginning of the thread.

Comments in the thread are created by a population of $U$ users, with each user producing a total number of comments equal to $\eta$, obeying the distribution $P(\eta)$. 
If a comment draws the attention of the OP, it can be rewarded with a positive acknowledgment. We refer to the authors of rewarded comments as \emph{receivers}, while we refer to the probability that users are rewarded by their \emph{fitness} $\rho$. We assume that the fitness $\rho$ is solely a function of the user's activity $\eta$: The more comments a user posts, the more likely the user is rewarded. For simplicity, we consider $\rho \sim \eta^\omega$, with $\omega > 0$. 
The probability that 
a randomly selected comment, written by a user with activity $\eta$, is rewarded is
\begin{equation}
p_r(\eta) = \frac{\eta^{\omega + 1}}{U \, n_0 \, \av{\eta^{\omega + 1}}} 
\; ,
\label{eq:prob1}
\end{equation}
where $n_0$ is the rate of rewarded comments and $\av{\ldots}$ denotes the mean value, see Appendix for details. 
Importantly, Eq.~\eqref{eq:prob1} is valid for a single comment. In case of $n(t)$ multiple comments posted on day $t$, the probability that 
at least one of the comments written by a user with activity $\eta$ is rewarded 
is given by $1 - \left[1 - p_r(\eta)\right]^{n(t)}$.
The goal of the model is to describe the attention dynamics of the OP, which is manifested by comment rewarding. 
More specifically, we are interested in finding the days $t = t^*$ in which comments are rewarded, assuming that rewards are assigned the same day the comments are created. 
We characterize the probability distribution that a comment is rewarded on a certain day $t^*$, $P(t^*)$. 
Such a probability is a function of the model's parameters. We assume that the rate of reward $n_0$ is constant in time. 
Also, we assume that $N$ and $T$ are not correlated and are generated from the distributions $P(N)$ and $P(T)$, respectively. 
For each thread, we know the function $n(t)$ describing the number of
comments made on day $t$; such a function is constrained by the actual size $N$ and duration $T$ of the thread. 
Finally, the population of users is fully described by the activity distribution $P(\eta)$ and the fitness exponent $\omega$. 
We first provide an analytical solution for $P(t^*)$ where the above quantities are given by mathematical functions; next, we validate our model using empirical distributions obtained from the CMV dataset.

\emph{Attention dynamics}. 
It has been widely observed that the collective attention received by academic articles, in the context of science, and cultural products, such as movies and songs, decays following an exponential function~\cite{parolo15, candia_universal_2019}. 
In line with these findings, we assume that the number of comments in threads decays in time according to the exponential function
\begin{equation}
    n(t) = A_{N,T} \, \exp{\left( -\lambda \frac{t}{T} \right)} 
    \; ,
\label{eq:fillcomments}
\end{equation}
where $t = 1, \ldots, T$, $\lambda$ is a tunable parameter, and $A_{N,T} = N \left[ \exp{(\lambda/T)} - 1 \right] / \left[ 1 - \exp{(-\lambda)} \right]$ ensures that the total number of comments over $T$ days is $N$.

\begin{figure}[tbp]
    \centering
    \includegraphics[width=0.9\columnwidth]{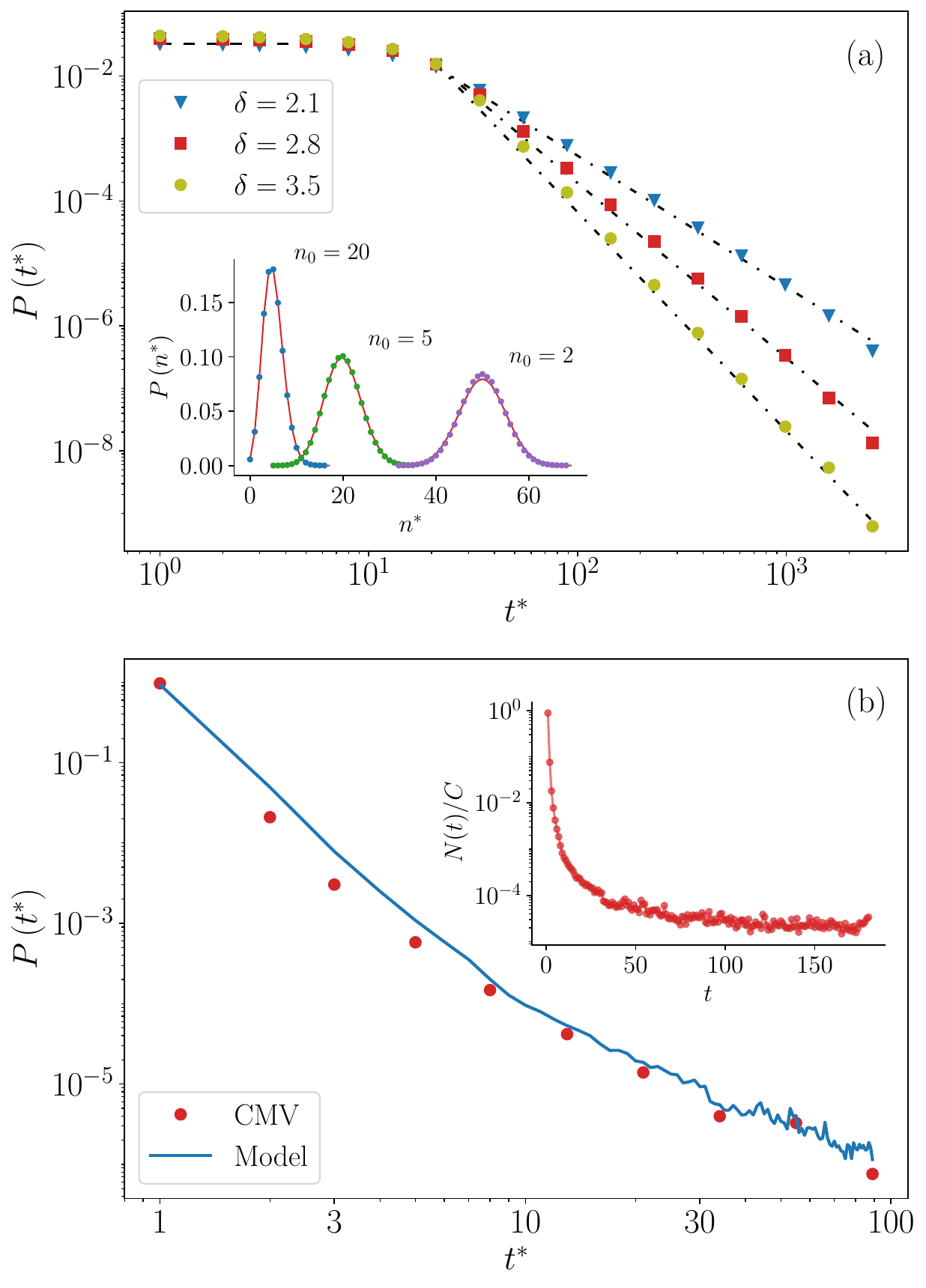}
    \caption{
    (a) Inset: Probability distribution of number of rewarded comments $P(n^*)$ in threads with $N = 100$ and $T = 33$, for different values of $n_0$. 
    Symbols correspond to numerical simulations, red lines correspond to the binomial distribution. 
    We used $\omega = 1$, $P(\eta) = 1$, $\lambda = 3$, and number of threads $V = 10^6$. 
    Main: Probability distribution of reward days $P(t^*)$ considering a power-law distribution $P(T)$, for different values of exponent $\delta$, see Eq.~\eqref{eq:T_powerlaw}. 
    Dashed lines following the predicted asymptotic behavior for $t^* \to 1$ ($P(t^*) \sim \mathrm{const.}$) and $t^* \to \infty$ ($P( t^*) \sim (t^*)^{-\delta}$) are included. 
    We used $n_0 = 100$, $\lambda = 3$, $T_0 = 30$, and number of threads $V = 10^6$. 
    (b) Inset: Normalized global time series $N(t)/C$ displayed in semi-log scale, representing the proportion of comments posted in CMV on day $t$.
    Main: Probability distribution of reward days $P(t^*)$ observed in CMV. Red symbols correspond to the real results, solid line corresponds to numerical results obtained from Eq.~\eqref{eq:numerical}.
    }
    \label{fig:solution}
\end{figure}

By averaging the probability of being rewarded in Eq.~\eqref{eq:prob1} over the users, one can see that the 
rate of reward is $n_0$, independent of the day $t$. 
Therefore, the probability that $n^*$ comments are rewarded in a thread follows a binomial distribution with number of trials equal to $N$ and the success probability equal to $1/n_0$, see inset of Fig.~\ref{fig:solution}(a).

For an exponential $n(t)$ given by Eq.~\eqref{eq:fillcomments}, in Appendix we show that $P(t^*)$ is fully determined by the threads' duration distribution $P(T)$ and is independent of the number of comments received, $P(N)$, and the activity of users, $P(\eta)$. 
As a particular case, we consider a power-law distribution $P(T)$ of the form
\begin{equation}
    P(T) = B \, T^{-\delta} \; ,
\label{eq:T_powerlaw}
\end{equation}
with $T \in [T_0, \infty)$, $B = (\delta - 1) / T_0^{-(\delta - 1)}$, and $\delta > 2$. 
For this form of the $P(T)$, 
we obtain that, for $t^* \to 1$, the distribution $P(t^*)$ goes to a constant 
if $\lambda \ll T_0$, while for $t^* \to \infty$ we get the asymptotic behavior $P(t^*) \sim (t^*)^{-\delta}$ (see Appendix for details). 
The main panel of Fig.~\ref{fig:solution}(a) shows the distribution $P(t^*)$ obtained from numerical simulations, which confirms the predicted behaviors. Importantly, to get a constant $P(t^*)$ for low values of $t^*$, we ensured that the condition $\lambda \ll T_0$ is satisfied. 
We observe that the crossover between the two predicted regimes---constant for $t^* \to 1$ and power-law for $t^* \to \infty$---occurs around $t^* = T_0$.


\emph{Data validation}. 
We empirically validate our model by measuring how individuals allocate their attention on CMV, where users reward their peers for interesting comments. 
On CMV, 
users submit posts expressing a certain opinion, and others respond to the post, attempting to change the OP's view by providing counterarguments. 
The OP acknowledges the relevant comments by rewarding the author of the successful comment. 
We use these rewards as traces to monitor the attention dynamics of the OP, i.e., to reconstruct the probability distribution $P(t^*)$ that a comment is rewarded at time $t^*$.

We collected data from CMV comprising $V = 102296$ threads and $C = 6564271$ comments posted by $U = 427742$ users from its creation in 2013 to 2023, see Section I of the Supplementary Material for details. Each thread $i$ is characterized by the number of comments $n_i(t)$ posted on day $t$ since the creation of the thread. The duration of the thread $i$ is denoted by $T_i$, which means that $n_i(t) = 0$ for $t > T_i$; the total size of the thread $i$ is then $N_i = \sum_t n_i(t)$. 
We empirically observe that $1 \leq T_i \leq 180$ for all threads, indicating that comments are not allowed more than 6 months after the post's creation date. 
We define the empirical distributions of thread duration as $P(T) = \frac{1}{V} \sum_{i=1}^V \delta_{T, T_i}$ and thread size as $P(N) = \frac{1}{V} \sum_{i=1}^V \delta_{N, N_i}$, where the Kronecker delta function is such that $\delta_{x,y} = 1$ if $x=y$ and $\delta_{x,y} = 0$ otherwise. 
Further, each user $u$ is characterized by the total number of comments posted $\eta_u$ and the total number of rewarded comments $0 \leq \eta^*_u \leq \eta_u$. The empirical distribution of users' activity is defined as $P(\eta) = \frac{1}{U} \sum_{u=1}^U \delta_{\eta, \eta_u}$.

We observe that the thread-specific time series $n_i(t)$ cannot be approximated by a simple analytical function. However, such information is not strictly required to model attention dynamics over all threads. To this end, we first create a global time series by aggregating all threads together, i.e., $N(t) = \sum_i n_i(t)$, see inset of Fig.~\ref{fig:solution}(b). 
Therefore, the total number of comments in CMV is $C = \sum_t N(t) = \sum_{t, i} n_i(t)$. 
We then keep explicit information about the duration and size of the $i$-th thread by defining the thread-specific time series
\begin{equation}
    \tilde{n}_i(t) = \frac{N_i\, N(t)}{\sum_{t'=1}^{T_i} N(t')} 
    \; .
\label{eq:n_general}
\end{equation}
As before, $\tilde{n}_i(t) = 0$ for $t > T_i$, so $\sum_t \tilde{n}_i(t) = N_i$. 
Notice that the previous equation can be seen as a generalization of Eq.~\eqref{eq:fillcomments} for $N(t) \sim e^{-\lambda t/T}$.

For the $i$-th thread, the conditional probability of observing a reward on day $t = t^*$, given $N_i$ and $T_i$, is defined as (see Appendix)
\begin{equation}
    P(t^* | N_i, T_i) = \frac{\tilde{n}_i(t^*)}{N_i} = \frac{N(t^*)}{\sum_{t' = 1}^{T_i} N(t')} \equiv P(t^* | T_i) 
    \; ,
\label{eq:prob_general}
\end{equation}
which is independent of the thread size $N_i$ and the activity of users $\eta$. 
By definition, we have 
$P(t^* | T_i < t^*) = 0$. 
The probability distribution $P(t^*)$ is obtained by averaging over all threads, thus
\begin{equation}
    P(t^*) = \sum_{i=1}^V P(t^* | T_i) = N(t^*) \sum_{i=1}^V \frac{1}{\sum_{t' = 1}^{T_i} N(t')} \; .
\label{eq:numerical}
\end{equation}
The main panel of Fig.~\ref{fig:solution}(b) shows the numerical distribution $P(t^*)$ obtained from Eq.~\eqref{eq:numerical}, represented by a solid line. We observe that the results of the model recover the real distribution obtained from CMV, represented by the symbols.

\emph{Inter-reward dynamics}. 
After investigating the probability that a comment is rewarded on a certain day, $P(t^*)$, we test if the model is able to capture the inter-reward dynamics. 
Even though $P(t^*)$ is independent of the users' fitness $\rho$, the fitness distribution $P(\rho)$ is central to determining the probability that a user is rewarded. 
We can measure the fitness of a user as the number of times that has been rewarded in total, in such a way that the fitness of receivers is nonzero. 
In Supplementary Fig.~1 we show the number of rewarded comments, $\eta^*$, as a function of the number of comments posted, $\eta$, for every receiver in CMV. 
As one could expect, there is a clear correlation between both quantities: The more (less) comments a user posts the more (less) rewards receives. 
However, since the dispersion is non-negligible, we cannot 
assume an analytical relationship of the form $\rho \sim \eta^\omega$
by directly performing a nonlinear regression. 
Instead, we can sample $\rho$ from the posterior distribution $P(\rho | \eta)$ obtained via the Bayes' rule~\cite{gelman2013bayesian}. 
First, we approximate the conditional probability $P(\eta | \rho)$ of observing $\eta$ given $\rho$ as normal distribution with mean $\mu$ and standard deviation $\sigma$ given by
\begin{equation}
    \mu(\rho) = \mathcal{C}_\mu \rho^{\nu_\mu} \; , \qquad
    \sigma(\rho) = \mathcal{C}_\sigma \rho^{\nu_\sigma} \; ,
\label{eq:normal_rho}
\end{equation}
with $\nu_\mu = 0.91$ and $\nu_\sigma = 0.68$, see regression analysis in Supplementary Fig.~2. Considering that the prior probability $P(\rho)$ is defined by the distribution of rewarded comments posted by a single user in CMV, $P(\eta^*) \sim (\eta^*)^{-\gamma_{\rho}}$, we can write the posterior distribution as
\begin{eqnarray}
    P(\rho | \eta) &=& \frac{P(\rho) \, P(\eta | \rho)}{P(\eta)} \nonumber \\
    &\sim& \rho^{-\left( \gamma_\rho + \nu_\sigma \right)} \exp{\left( -\frac{\left[ \eta - \mu(\rho) \right]^2}{2 \sigma^2(\rho)} \right)} 
    \; .
\label{eq:bayes_attr}
\end{eqnarray}

\begin{figure}[tbp]
    \centering
    \includegraphics[width=0.9\columnwidth]{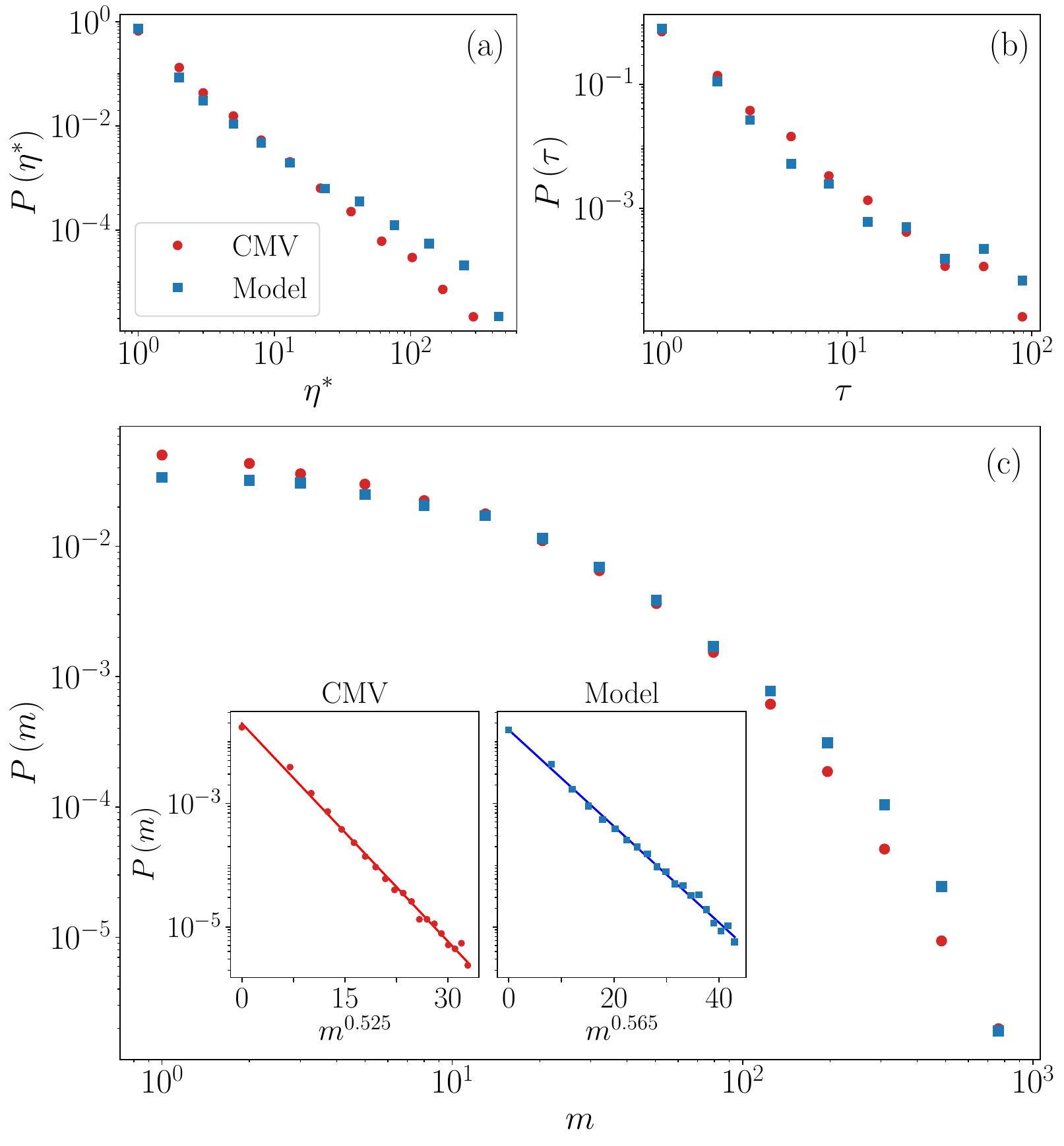}
    \caption{
    (a) Probability distribution $P(\eta^*)$ of number of comments posted by a single user that have been rewarded. 
    (b) Probability distribution $P(\tau)$ of elapsed days between two consecutive rewarded comments in threads. 
    (c) Main: Probability distribution $P(m)$ of elapsed comments between two consecutive rewards of a single user. 
    Insets: Nonlinear regression of a stretched exponential performed in $P(m)$ with exponents $\beta = 0.52 \pm 0.02$ (CMV) and $\beta = 0.56 \pm 0.02$ (model). 
    We show the results obtained from both CMV and the model, represented by red and blue symbols, respectively. 
    Very similar results are obtained by running the model with different initial conditions. 
    We used a logarithmic binning where bins are sized like the Fibonacci sequence~\cite{vigna2013fibonacci}. 
    We used $n_0 = 200$ in Eq.~\eqref{eq:prob1}.
    }
    \label{fig:CMVmodel}
\end{figure}

We feed our mathematical model using spline interpolations~\cite{2020SciPy-NMeth} of the empirical distributions $P(N)$, $P(T)$, and $P(\eta)$, see Supplementary Fig.~3. 
Similarly, we use the thread-specific time series given by Eq.~\eqref{eq:n_general} as the number of comments throughout the days. 
In Fig.~\ref{fig:CMVmodel}(a) we compare the distribution of rewarded comments $P(\eta^*)$ obtained in the model with the one directly measured in the data. 
We observe that the model reproduces the real distribution with good accuracy, showing a power-law signature $P(\eta^*) \sim (\eta^*)^{-\gamma_{\rho}}$ with exponents $\gamma_{\rho} = 2.19 \pm 0.02$ in the data and $\gamma_{\rho} = 1.86 \pm 0.07$ in the model. 
It should be stressed that, according to the empirical data, we restrict the possible number of rewards per user in the same thread to 1. This means that if a receiver has 100 rewards in CMV, they must be rewarded in 100 different threads.

We also study the inter-reward dynamics by counting the number of days between two consecutive rewarded comments. 
By definition, the first possible reward in a thread is given in $t = 1$. 
Therefore, if, for example, the first rewarded comment is posted on day $t = 1$, the number of elapsed days is $\tau = 0$. If the second rewarded comment is posted on $t = 5$, then $\tau = 4$. 
We compare the distribution of elapsed days $P(\tau)$ observed empirically in CMV with the one obtained by numerical simulations of the model. 
To this aim, in the model we consider $V = 10^5$ threads and $U = 4.5 \times 10^5$ users. 
Figure~\ref{fig:CMVmodel}(b) shows the interevent time distribution $P(\tau)$ in CMV (red symbols) and in the model (blue symbols). 
We observe that the model reproduces the real distribution with good accuracy, showing a power-law signature $P(\tau) \sim \tau^{-\gamma_t}$ with exponents $\gamma_t = 2.30 \pm 0.08$ in the data and $\gamma_t = 2.02 \pm 0.17$ in the model.


Finally, we study the inter-reward dynamics from the point of view of the 
commenters. 
For each receiver, i.e., a user who has been rewarded at least once, we define the time-ordered series of comments they authored on different threads. 
Then, we express the elapsed time between two consecutive rewards as the number of comments occurred between them. 
Again, the first possible reward is given in the first position in the series. 
Therefore, if, for example, the first reward is given to the first comment, the number of elapsed comments is $m = 0$. 
If the second reward is given to the fifth comment, then $m = 4$. 
The main panel of Fig.~\ref{fig:CMVmodel}(c) shows the distribution of elapsed comments $P(m)$ observed empirically in CMV (red symbols) and obtained numerically from the model (blue symbols). 
We observe that the model accurately reproduces the real distribution. 
In the insets of Fig.~\ref{fig:CMVmodel}(c), we show that the distribution $P(m)$ follows a stretched exponential $P(m) \sim \exp{\left( -a \, m^{\beta} \right)}$. 
By performing a nonlinear regression (solid lines), we obtain stretching exponents $\beta = 0.52 \pm 0.02$ in the data and $\beta = 0.56 \pm 0.02$ in the model.

\emph{Discussion}. 
In this work, we addressed a key gap in the study of attention dynamics by shifting the focus from collective behavior to an individual level. 
While much of the previous research has examined how information spreads, trends emerge, or memes compete at large scale, we focus on a microscopic perspective: 
How a single user allocates their attention in the face of many competing stimuli. 

Our main contribution is a minimal, analytically tractable model that captures how individual attention decays over time, as a function of the total duration of interactions, rather than their number or the activity levels of interacting users. 
We validated the model using empirical data from the subreddit \textit{Change My View}, where OPs engage with comments by rewarding them---an observable proxy for individual attention. 
Despite its simplicity, the model successfully reproduced both the decay in OP's attention and the interevent times between successive rewards. 
These findings underscore the model’s predictive power and its potential applicability to a wide range of online contexts, from email management to collaborative work platforms.


The model's simplicity trades off realism for analytical tractability, which opens several avenues for future research. 
From a theoretical point of view, the model proposed can be interpreted as a stochastic counting process of attention allocation, susceptible to an application of the tools of renewal theory~\cite{cox_renewal_1967} to obtain further information. 
From the methodological side, on the other hand, heterogeneity could be incorporated in user fitness, linking it not only to activity but also to an intrinsic capacity to steer focal user's attention, content-based preference, or interaction similarity between users. 
In this setting, the individual attention dynamics is likely to depend also on the heterogeneity of the fitness distribution. 
Furthermore, one could validate our results on other data sets describing distinct social contexts, where the individual attention dynamics may be driven by different mechanisms. 
The generality of the model can be tested by considering alternative online platforms different from Reddit. 
Finally, it has been recently shown that individual attention is closely related to how users express their emotions on social media~\cite{zhao_attention_2025}. Similarly to our work, future efforts should extend the framework of collective emotions to a microscopic, individual level~\cite{schweitzer_agent-based_2010, chmiel_collective_2011}.


\emph{Acknowledgments}. 
J.~O., S.~F., and M.~S. acknowledge the support of the AccelNet-MultiNet program, a project of the National Science Foundation (award \#1927425 and \#1927418). 
J.~O. and R.~P.-S. acknowledge financial support from project PID2022-137505NB-C21/AEI/10.13039/501100011033/FEDER UE. 
F.~R. acknowledges support from the Air Force Office of Scientific Research under grant number FA9550-24-1-0039. 
M.~S. acknowledges support from Grants No. RYC2022-037932-I and CNS2023-144156 funded by MCIN/AEI/10.13039/501100011033 and the European Union NextGenerationEU/PRTR. 
We thank the Indiana University Observatory on Social Media (OSoMe) for managing and providing access to Reddit data. We also acknowledge Jisun An and Haewoon Kwak for initially bringing this dataset to OSoMe's attention.

\emph{Data availability}. 
The Reddit data was originally compiled by Pushshift and \texttt{u/Watchful1} and subsequently updated by Arthur Heitmann as part of the Arctic Shift project (\url{https://github.com/ArthurHeitmann/arctic_shift}). The version used in this study was accessed via Academic Torrents (\url{https://academictorrents.com/details/56aa49f9653ba545f48df2e33679f014d2829c10}) in February 2024.




\section*{Appendix A: Solution of the model}

Comments are created by a population of $U$ users. Focus on one randomly selected user which we indicate as the focal user. We denote by 
\[
P(c=1) = \frac{1}{U}
\]
the probability that a randomly selected comment is authored by the focal user, and by
\[
P(r=1) = \frac{1}{U} \frac{R}{N} = \frac{1}{U \, n_0}
\]
the probability that a randomly selected comment that is rewarded is authored by the focal user. In the above expression, $n_0 = N/R$ is the reward rate, that is, the ratio between the total number of comments $N$ and the total number of rewarded comments $R$.

The population of users is characterized by the distribution $P(\eta)$, with $\eta$ representing the total number of comments posted by a user, or activity. Overall user activity and comments are related by
\[
U \, \av{\eta} = N \; ,
\]
where
\[
\av{\eta^k} = \sum_\eta P(\eta) \, \eta^k
\]
is the $k$-th moment of the distribution $P(\eta)$.

Focus now on a randomly selected user with activity $\eta$. 
The probability that a randomly selected comment is generated by the focal user with activity $\eta$ is $P(\eta | c=1) \sim \eta P(\eta)$, thus
\[
P(\eta | c=1) = \frac{\eta P(\eta)}{\av{\eta}}
\]
upon proper normalization. 
The probability that the focal user with activity $\eta$ is the author of a randomly selected comment is
\[
P(c=1 | \eta) = \frac{P(\eta | c=1) \, P(c=1)}{P(\eta) } = \frac{\eta}{U \, \av{\eta}} \; .
\]

To model the dependence between the mechanism that drives comment rewarding and the activity of users, we assume that the probability that a rewarded comment is authored by the focal user with activity $\eta$ is
\[
P(\eta | r=1) \sim \eta^\omega \, P(\eta | c=1) \; ,
\]
where the term $\eta^{\omega}$ reflects the activity-dependent likelihood of a comment written by a user with activity $\eta$ to receive a reward. We can write that
\[
P(\eta | r=1) = \frac{\eta^{\omega+1}}{\av{\eta^{\omega+1}}} \, P(\eta) \; ,
\]
upon proper normalization. In turn, the probability that 
the focal user with activity $\eta$ is the author of a randomly selected comment that is rewarded is (see Eq.~\eqref{eq:prob1})
\[
P(r=1 | \eta) = \frac{P(\eta | r=1) \, P(r=1)}{P(\eta)} 
= \frac{\eta^{\omega+1}}{U \, n_0 \, \av{\eta^{\omega+1}}} 
\; .
\]
The above expression is valid for a single comment. If we consider $N$ comments, the probability that the focal user does not receive a reward in any of these $N$ comments is $\left[1 - P(r=1| \eta) \right]^{N}$.

We now turn our attention to the dynamical process of receiving rewards in a thread of size $N$ and duration $T$. We assume that the rate of reward is constant in time, i.e., $n_0 = n(t)/r(t)$ with $n(t)$ and $r(t)$ the total number of comments and rewarded comments in a thread with age equal to $t$, respectively. By definition, $n(t) = 0$ for $t > T$ and $\sum_{t} n(t) = N$. 
The probability that 
the focal user with activity $\eta$ receives a reward for the first time in the thread at $t = t^*$ is therefore
\[
P(t=t^* | \eta, N, T) = \frac{Q(t^*-1) - Q(t^*)}{1 - Q(T)} \; ,
\]
where
\[
Q(t) = \prod_{t'=1}^t \, \left[1 - P(r=1 | \eta) \right]^{n(t')} \; .
\]
Assuming that $\frac{\eta^{\omega+1}}{U n_0 \av{\eta^{\omega+1}}} \ll 1$, we can approximate
\[
Q(t) \simeq 
1 - \frac{\eta^{\omega+1}}{U \, n_0 \, \av{\eta^{\omega+1}}}  \sum_{t'=1}^t n(t') \; .
\]
Thus, we find that
\[
Q(t^*-1) - Q(t^*) = \frac{\eta^{\omega+1}}{U \, n_0 \, \av{\eta^{\omega+1}}} \, n(t^*) \; .
\]
Also, we clearly have that
\[
1 - Q(T) = \frac{\eta^{\omega+1}}{U \, n_0 \, \av{\eta^{\omega+1}}} \, N \; .
\]
In summary,
\begin{equation}
P(t=t^* | \eta, N, T) = \frac{n(t^*)}{N} \equiv P(t^*|N, T) \; ,
\label{eq:prob_moregeneral}
\end{equation}
which is independent of the activity $\eta$ of the focal user.

If the number of comments $n(t)$ posted on day $t$ is given by the exponential function of Eq.~\eqref{eq:fillcomments}, the conditional probability Eq.~\eqref{eq:prob_moregeneral} is
\[
P(t^* | N, T) = \frac{\exp{\left( -\lambda \frac{t^* - 1}{T} \right)} - \exp{\left( -\lambda \frac{t^*}{T} \right)}}{1 - \exp{( -\lambda)}} \equiv P(t^* | T) \; ,
\]
which is independent of $N$ and 
is normalized for discrete duration $T$. In the continuum limit with $T \in [T_0, \infty)$, the distribution should be normalized as
\begin{eqnarray}
    p(t^* | T) &=& \frac{P(t^* | T)}{\int_1^T dt' \, P(t' | T)} \nonumber \\ 
    &=& \frac{\lambda \left[ \exp{\left( -\lambda \frac{t^* - 1}{T} \right)} - \exp{\left( -\lambda \frac{t^*}{T} \right)} \right]}{T \left[ 1 - \exp{\left( -\frac{\lambda}{T} \right)} -  \exp{\left( -\lambda \frac{T - 1}{T} \right)} + \exp{(-\lambda)} \right]} \nonumber \\
    &\simeq& \exp{\left( -\lambda \frac{t^* - 1}{T} \right)} - \exp{\left( -\lambda \frac{t^*}{T} \right)} \; , \nonumber
\end{eqnarray}
where we assumed $T$ large. 
Therefore, averaging over $T$ we have
\[
P(t^*) = \int_{T_0}^\infty dT \, p(t^* | T) \, P(T) \; ,
\]
where $P(T)$ is the probability density function of $T$. 
Considering a power-law distribution (see Eq.~\eqref{eq:T_powerlaw})
\[
P(T) = \frac{\delta - 1}{T_0^{-(\delta - 1)}} \, T^{-\delta}
\]
with $\delta > 2$, 
$P(t^*)$ can be analytically calculated, yielding
\begin{eqnarray}
    P(t^*) &=& D \, (t^* - 1)^{-(\delta - 1)} \, \gamma \left( \delta - 1, \, \lambda \frac{t^* - 1}{T_0} \right) \nonumber \\ 
    &-& D \, (t^*)^{-(\delta - 1)} \, \gamma \left( \delta - 1, \, \lambda \frac{t^*}{T_0} \right) \; ,
\label{eq:prob_result}
\end{eqnarray}
where $D = (\delta - 1) (\lambda/T_0)^{-(\delta - 1)}$ and $\gamma(s, x)$ is the lower incomplete gamma function~\cite{abramovitz}. 
For extreme values of $t^*$, we can use the asymptotic expansions of $\gamma(s,x)$:
\begin{itemize}
    \item $\bm{t^* \to 1}$. Since $\gamma(s, x) \to x^s/s$ as $x \to 0$, the first term of Eq.~\eqref{eq:prob_result} tends to a constant. The same applies for the second term if and only if $\lambda \ll T_0$. 
    However, the limit in the first term converges faster than in the second term. Therefore, $P(t^*)$ tends to a positive constant that depends weakly on exponent $\delta$.
    \item $\bm{t^* \to \infty}$. Since $\gamma(s, x) \to \Gamma(s)$ as $x \to \infty$, where $\Gamma(s)$ is the gamma function, we have
    \begin{eqnarray}
        P(t^*) &\sim& (t^* - 1)^{-(\delta - 1)} - (t^*)^{-(\delta - 1)} \nonumber \\ 
        &=& t^* (t^* - 1)^{-\delta} - (t^* - 1)^{-\delta} - t^* (t^*)^{-\delta} \nonumber \\
        &\simeq& (\delta - 1) (t^*)^{-\delta} - \delta (t^*)^{-(\delta + 1)} \nonumber \\
        &\sim& (t^*)^{-\delta} \; . \nonumber
    \end{eqnarray}
\end{itemize}

\bibliography{cmv_reddit}

\clearpage

\newcommand{\thisisthetitle}{Modeling individual attention dynamics on online social media}

\pagestyle{fancy}

\lhead{\thisisthetitle\ - SM}
\chead{}
\rhead{}

\lfoot{J. Ojer, F. Radicchi, S. Fortunato, M. Starnini and R. Pastor-Satorras}
\cfoot{}
\rfoot{\thepage}

\renewcommand{\headrulewidth}{0.5pt}
\renewcommand{\footrulewidth}{0.5pt}

\setcounter{section}{0}
\renewcommand{\thefigure}{\textbf{SF~\arabic{figure}}}
\setcounter{figure}{0}
\renewcommand{\figurename}{\textbf{Supplementary Figure}}
\setcounter{equation}{0}

\title{\thisisthetitle\ \\ \ \\
  Supplementary Material}

\author{Jaume Ojer}
\affiliation{Departament de F\'isica, Universitat Polit\`ecnica de Catalunya, Campus Nord, 08034 Barcelona, Spain}

\author{Filippo Radicchi}
\affiliation{Center for Complex Networks and Systems Research, Luddy School of Informatics, Computing, and Engineering, Indiana University, Bloomington, Indiana 47408, USA}

\author{Santo Fortunato}
\affiliation{Center for Complex Networks and Systems Research, Luddy School of Informatics, Computing, and Engineering, Indiana University, Bloomington, Indiana 47408, USA}

\author{Michele Starnini}
\affiliation{Department of Engineering, Universitat Pompeu Fabra, 08018 Barcelona, Spain}
\affiliation{CENTAI Institute, 10138 Turin, Italy}

\author{Romualdo Pastor-Satorras}
\affiliation{Departament de F\'isica, Universitat Polit\`ecnica de Catalunya, Campus Nord, 08034 Barcelona, Spain}

\maketitle
\onecolumngrid
\begin{center}
\textbf{\large Supplementary Material}
\end{center}

\section{Empirical analysis of CMV}

Reddit is structured around subject-specific communities known as subreddits, each dedicated to a particular topic or interest. Within these subreddits, individuals can post content and respond to both original posts and other comments, forming a branching discussion tree. 
One such community, \texttt{r/changemyview} (CMV)---the focus of this study---encourages users to engage in discussions aimed at exploring and challenging differing opinions and perspectives revolving around political and socioeconomic issues. 
In particular, the original posters (OPs) express their opinion regarding any topic, and commenters respond to the post by attempting to change their view. If the OP's opinion is changed, they can reward successful commenters with a Delta symbol $\Delta$.

We analyze CMV data over 10 years, from 2013 to 2023. The total number of threads and comments are $V' = 255287$ and $C' = 11461626$, respectively. 
However, these interactions are sometimes authored by users whose Reddit account has been deleted or removed. Additionally, users can be bots that imitate human activity. All threads and comments generated by deleted users and bots are discarded. 
In order to identify bot users, we relied on public lists enumerating known bots in Reddit, such as B0tRank (\url{botrank.pastimes.eu}) and reported bots in \texttt{r/BotDefense}. We also detected potential bots by analyzing the text in comments, such as repetition of words and constant use of very short sentences. Furthermore, we excluded users that have the string 'bot' in their username, as well as a very large activity in Reddit, like posting more than 50000 comments or interacting in more than 1000 different subreddits. 
Finally, the OP usually replies to all comments, while just a few draw the attention of the OP, change their view, and are rewarded with a $\Delta$. Therefore, in order to focus on comment rewarding, we only take into account comments posted by users different from the OP.
After discarding deleted users, bots, 
and OP's comments,
the number of threads is $V = 102296$ containing $C = 6564271$ comments posted by $U = 427742$ users.

\clearpage

\section{Supplementary Figures}

\begin{figure}[h!]
    \centering
    \includegraphics[width=0.6\columnwidth]{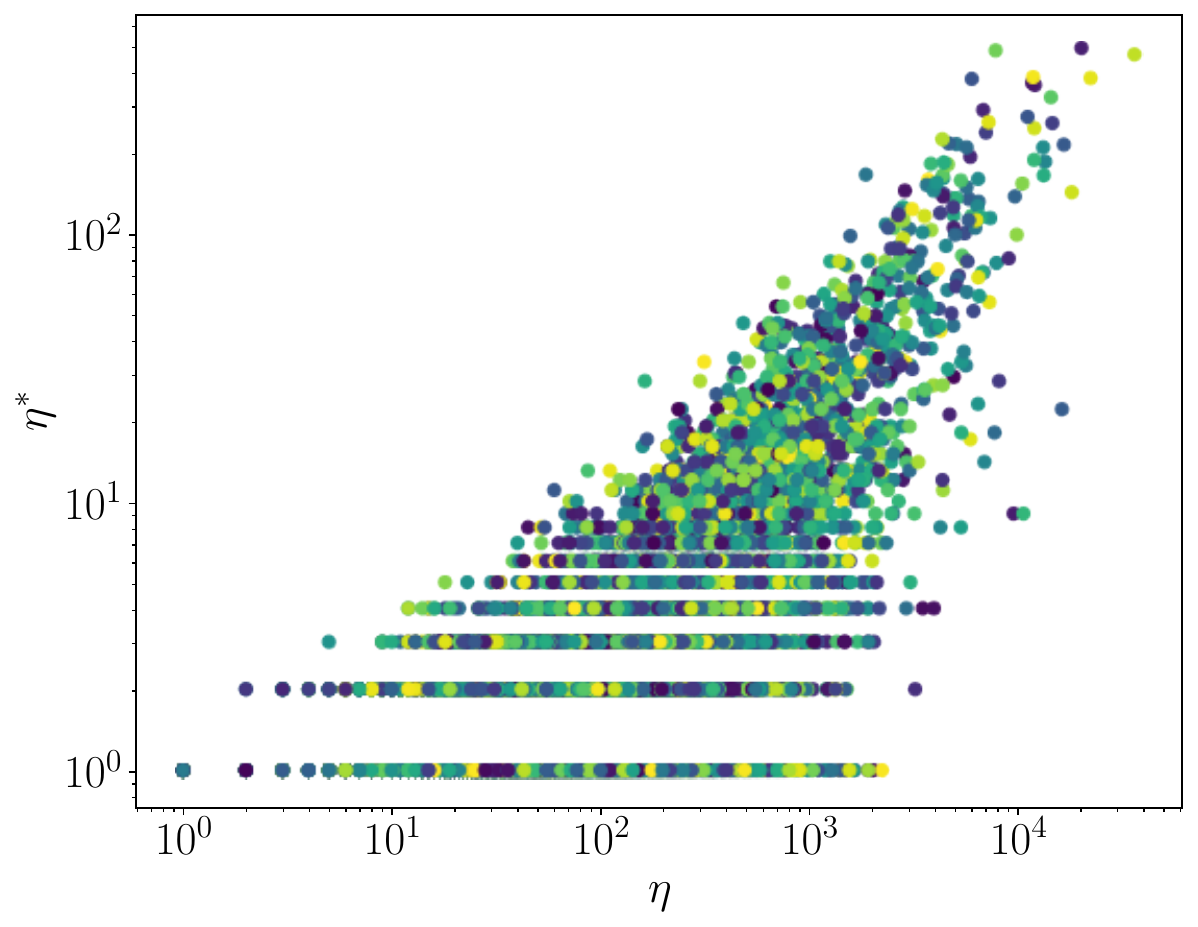}
    \caption{
    Number of rewarded comments as a function of the number of comments posted for every receiver in CMV. Every point depicted in a random color corresponds to a different user.
    }
    \label{fig:attractiveness}
\end{figure}

\begin{figure}[h!]
    \centering
    \includegraphics[width=0.9\columnwidth]{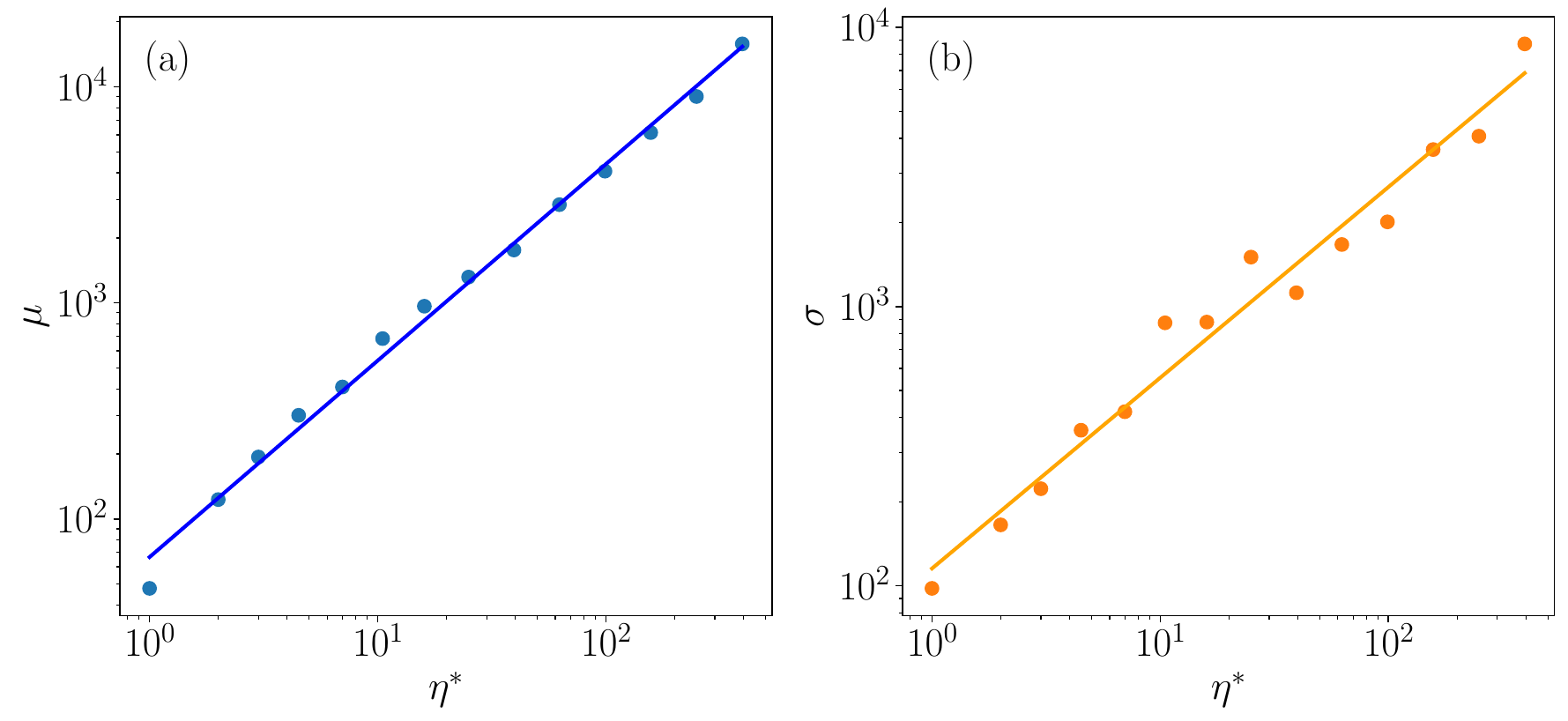}
    \caption{
    Mean value (a) and standard deviation (b) of the number of comments posted by receivers in CMV as a function of the number of rewarded comments. 
    Solid lines correspond to linear regressions in logarithmic scale: $\mu = \mathcal{C}_\mu (\eta^*)^{\nu_\mu}$ with $\mathcal{C}_\mu = 66.51 \pm 1.07$ and $\nu_\mu = 0.91 \pm 0.02$, $\sigma = \mathcal{C}_\sigma (\eta^*)^{\nu_\sigma}$ with $\mathcal{C}_\sigma = 115.28 \pm 1.13$ and $\nu_\sigma = 0.68 \pm 0.03$.
    }
    \label{fig:musigma}
\end{figure}

\begin{figure}[h!]
    \centering
    \includegraphics[width=0.95\columnwidth]{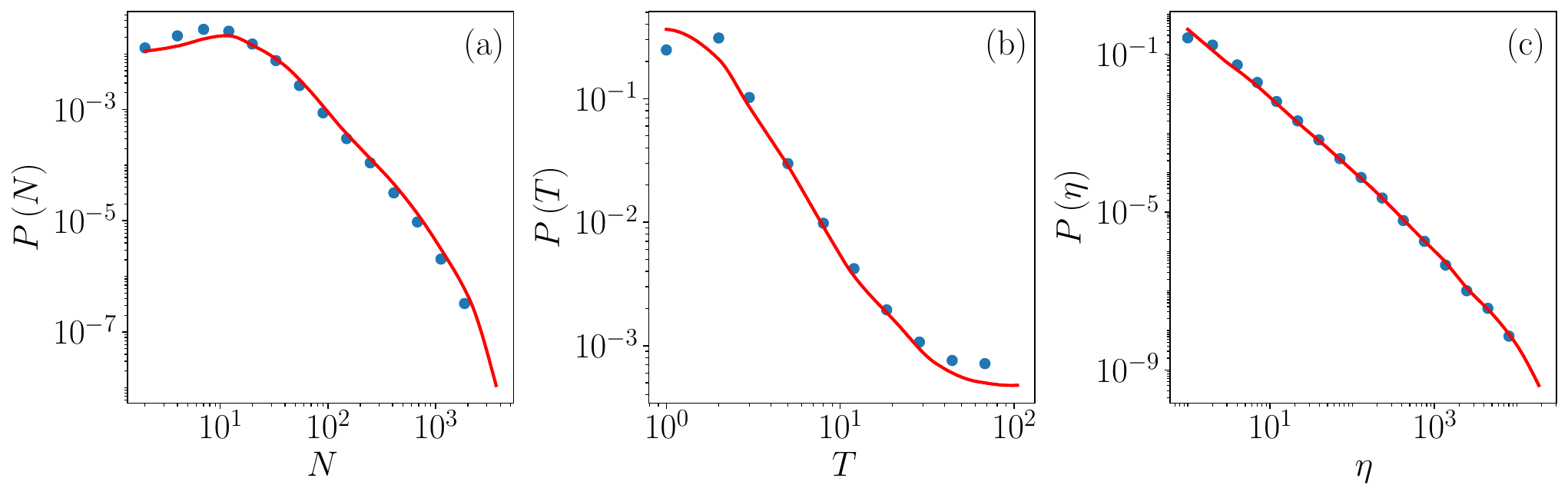}
    \caption{
    Empirical probability distributions of (a) total comments of threads $P(N)$, (b) duration of threads $P(T)$, and (c) activity of users $P(\eta)$ in CMV. 
    Red lines correspond to spline interpolations, blue symbols correspond to rejection sampling generation.
    }
    \label{fig:splines}
\end{figure}

\begin{figure}[h!]
    \centering
    \includegraphics[width=0.6\columnwidth]{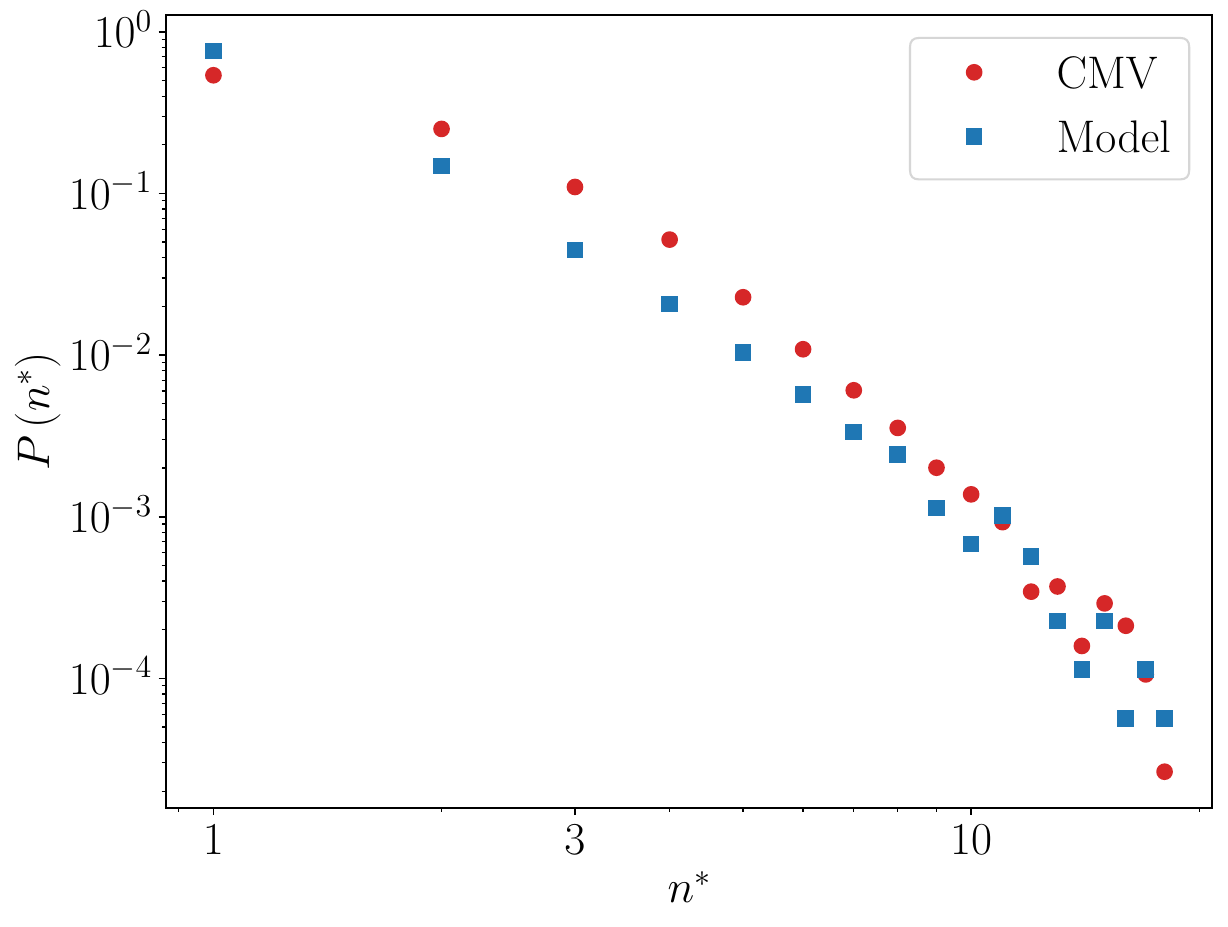}
    \caption{
    Probability distribution of number of rewarded comments $P(n^*)$ in threads from CMV and the model, represented by red and blue symbols respectively. 
    The distribution follows a stretched exponential with exponent $\beta = 0.56$ in the data and $\beta = 0.31$ in the model. 
    }
    \label{fig:PnDelta}
\end{figure}


\end{document}